\documentstyle[12pt,aasms4]{article}
\begin{document}

\title{Periodic Optical Outbursts from the Be/Neutron Star Binary
AX~J0049.4$-$7323\footnote{This paper utilizes public domain data obtained
by the MACHO Project, jointly funded by the US Department of Energy
through the University of California, Lawrence Livermore National
Laboratory under contract No.\ W-7405-Eng-48, by the National Science
Foundation through the Center for Particle Astrophysics of the University
of California under coopertative agreement AST-8809616, and by the Mount
Stromlo and Siding Spring Observatory, part of the Australian National
University. } } 

\author{A.P. Cowley \& P.C. Schmidtke}  
\affil{Department of Physics \& Astronomy, Arizona State University,
Tempe, AZ, 85287-1504 \\ email: anne.cowley@asu.edu}

\begin{abstract}

The optical light curve of the Be/neutron star binary AX~J0049.4$-$7323 has
been investigated using data from the MACHO and OGLE-II projects.  This
X-ray source, whose neutron star has a very slow rotation rate
(P$_{pulse}$= 755.5 sec), shows optical outbursts every $\sim$394 days. 
The regularity of these outbursts suggests that their recurrence time is
the orbital period of the system.  During the outbursts the system
brightens and becomes slightly redder.  A possible interpretation is that 
a portion of the equatorial disk is excited as the neutron star passes through 
it during periastron passage.  In the intervals between outbursts the light
curve shows $\sim$11-day quasi-periodic varability which may be associated
with the rotation of the Be star's extended disk. 
  
\end{abstract}


\section{Introduction}

Numerous X-ray pulsars have been found in the Magellanic Clouds, many of
which are associated with binary systems containing a Be star and a
neutron star (e.g. Yokogawa et al. 2000, Yokogawa et al. 2003, Haberl \&
Sasaki 2000, and references therein).  Corbet (1984) presented evidence
that a relationship exists between the neutron star spin period and the
orbital period in these systems.  This was further substantiated by Waters
\& van Kerkwijk (1989) who explained the correlation in terms of the
equilibrium spin period of the pulsar as a result of the slow equatorial
winds from the Be star.  Subsequent data have further strengthened this
spin/orbit correlation (e.g. see Fig.\ 6 by in't Zand et al. 2001). 

ASCA observations by Yokogawa et al.\ (2000) showed that AX~J0049.4$-$7323
has an unusually long pulse period (755.5 s), making it the slowest known
rotator among the X-ray pulsars in the SMC.  Hence, one expects that the 
system to have a long orbital period, based on Corbet's results from other 
Be/neutron star systems.  Using photometric data from the MACHO and OGLE-II 
Projects, we have identified what we believe to be the orbital period of 
$\sim$394 days for AX~J0049.4$-$7323, based on recurrent optical outbursts. 

From optical observations, Edge \& Coe (2003) identified
AX~J0049.4$-$7323 with a Be star which showed a double peaked H$\alpha$
emission profile at the time of their observations.  They give the
magnitude and color of the system as $V$=14.99 and $B-V$=0.05, but they
did not address the question of the orbital period or light variability. 
Using MACHO and OGLE-II data we have investigated the light curve of this 
Be star and found what we propose to be the orbital period as well as
quasi-periodic variability between outbursts. 

\section{Analysis of Optical Data from the MACHO and OGLE-II Projects}

The interactive MACHO Project light curve browser\footnote{
http://www.macho.mcmaster.ca/Data/MachoData.html} was used to extract
instrumental magnitudes for AX~J0049.4$-$7323.  Data on this system are
available for $\sim$7.4 years from MJD 48855 through MJD 51546.  The MACHO
website displays light curves called ``blue" and ``red," but after using
of the transformation equations given by Alcock et al.\ (1999), the colors
are close to $V$ and $R$, respectively.  Here, we refer to these bands as
$V_{MAC}$ and $R_{MAC}$, as we have done in previous papers using MACHO
data. 

$I$ magnitudes for this system are also available in the OGLE-II
database\footnote{http://bulge.princeton.edu/$\sim$ogle/dia/} where this
object is designated as OGLE004942.00$-$732314.2 in the ``difference
image analysis" of SMC variable stars.  $I$ magnitudes are given for dates
from 1997 January through 2000 July (MJD 50466 -- 51756), with the data
partially overlapping in time the last portion of the MACHO observations. 
  
\subsection{The 394-day Outburst Cycle} 

In Figure 1 we plot the MACHO and OGLE-II light curves for
AX~J0049.4$-$7323 in $V_{MAC}$, $R_{MAC}$, and $I$.  It is clear,
especially in the $R_{MAC}$ data, that outbursts up to $\sim$0.3 mag occur
regularly.  The outbursts have been labeled A through F, so that it is
possible to refer to them in the text.  The amplitude of the outbursts
varies, and since their duration is short, not all outbursts have
necessarily been observed at maximum light.  In the OGLE-II data, only the
last three outbursts were covered, and outburst D was first observed after
its start in the other two colors, so its time of maximum is not well
determined. 

Figure 2 displays an expanded light curve of outburst F in $V_{MAC}$,
$R_{MAC}$, $I$, and the color curve, $V_{MAC}-R_{MAC}$.  In all colors the
outburst shows a rapid rise and a more gradual decline.  The amplitude is
greater at longer wavelengths, implying the system as a whole reddens
during the outburst, as is shown in the bottom panel of this figure.  In
Figure 3 we plot $R_{MAC}$ light curves and $V_{MAC}-R_{MAC}$ color curves
for outbursts B through E.  One can see that both the amplitude and
duration of the outbursts vary.  During most outbursts the color of the
system reddens slightly, suggesting that the brightened region was cooler
than the Be star itself.  Typically, the equatorial disks surrounding Be
stars are cooler than the stars themselves (McGowan \& Charles 2002), so
we expect that a region in the disk has brightened. 
  
Using the MACHO data for outbursts C through F, we find the recurrence
time to be 394.3$\pm$2.3 days.  Outburst A is questionable, since it was only
observed by a single point in $R_{MAC}$ (with a large error) and missed in
$V_{MAC}$.  Using outbursts A through F in $R_{MAC}$, we find a period of
393.0 days.  Similarly, from $V_{MAC}$ data for outbursts B through F we 
derive a period of 394.5 days.  We adopt 394 days as a mean period, noting
that for some outbursts the time of peak brightness was not well
determined.  The OGLE-II data are in good agreement with this value, even
though only one outburst (F) was fully observed. 

The regularity of these outbursts suggests that this 394-day period is
likely to be the orbital period of AX~J0049.4$-$7323.  Placing this system
on the ``Corbet diagram" (pulse period vs.\ orbital period) given in Fig.\
6 by in't Zand et al.\ (2001), we see a very good fit at the extreme upper
end of the relationship for Be-star X-ray systems.  Hence, 394 days is
very likely the orbital period.  Since these systems typically have
eccentric orbits, a possible interpretation is that the outbursts occur
when the neutron star enters the extended equatorial disk (at or near
periastron passage) surrounding the Be star.  Variations in the amplitude
and onset of the outbursts may depend on the density and extent of the
disk at the time the neutron star plunges into the disk. 

Similar outburst behavior is already known in several other Be/neutron
star binaries.  The most noted case is A0538$-$66 in the LMC, whose
orbital eccentricity in its 16.6-day period is estimated to be $e\sim0.7$
(e.g. Charles et al. 1983, Hutchings et al. 1985).  During periastron
passage, this system brightens by as much as 2 mag, although there is
considerable variation in amplitude of the outbursts (e.g. Alcock et al.
2001).  Several other Be/neutron star binaries are known to show smaller
periodic outbursts which are thought to occur near periastron, although
there have been no direct measurements of their orbital eccentricities.
These include RX~J0520.5$-$6921 (Coe et al. 2001) and RX~J0058.2$-$7231
(Schmidtke et al. 2003), with orbital periods near 24 and 60 days,
respectively.  Both show outburst amplitudes of $\sim$0.05 mag as well as
reddening of the system during the outburst, as is seen here for
AX~J0049.4$-$7323. 

\subsection{Quasi-Periodic Variability between Outbursts}

In addition to the regular outburst cycle, we have examined the photometry
between outbursts to see if there was any periodic behavior.  We first looked 
at Segment 3 which shows a relatively constant mean magnitude (see
Fig.\ 1).  We pre-whitened the data by subtracting a linear trend to produce 
residuals designated as $R^\ast$.  Using the method described by Horne \& 
Baliunas (1986), we carried out an analysis of the power spectrum over a 
large range of frequencies.
There is a relatively strong signal at P=10.69 days 
(labeled `f') and a weaker one at its alias periods,  1.1 days (labeled 
`1$-$f').  A similar analysis of $V_{MAC}$ data for the same time span yields 
a primary period at 10.61 days.  Both colors show a sinusoidal light curve 
with a full amplitude of $\sim$0.03 mag during this segment.  

We then analyzed the other numbered segments in all available colors.  Each
segment was detrended using either a linear or quadratic fit prior to
calculating the periodograms.  Strong peaks are present in the 
periodograms for Segments 2, 4, and 5, with mean periods of 10.89, 10.86, 
and 11.35 days, respectively.  The power spectra for Segments 2 -- 5 are shown 
in the bottom panel of Figure 4.   One can see that the periods with the 
most power in Segments 2 and 4 differ from those found for Segments 3 and 5. 
Even for Segment 1, which shows an overall downward trend in brightness, 
there is a weak peak at 10.37 days in the $R^\ast$ data.  Segment 6, which 
is only well represented in $I$, shows a peak at 11.42 days. 

In Figure 5 we plot the phased $R^\ast$ light curves for Segments 2 and 3, 
folded on their individual 11-day periods.  Although the amplitude of the 
folded curves varies substantially between segments, there is no dependence 
on either period or mean brightness. 

We also searched for coherent changes in $V_{MAC}-R_{MAC}$ color at
periods similar to those present in the individual bandpasses, but found
none.  Except for the reddening during outbursts, the system's color 
remained remarkably constant after MJD 49600.  The mean value ranged from
$V_{MAC}-R_{MAC}$ = 0.14$\pm$0.03 in Segment 2 to 0.15$\pm$0.03 in
Segment 4.  Prior to MJD 49600, however, small changes in color were
observed.  The system was bluest at the beginning of the MACHO observing,
with $V_{MAC}-R_{MAC} \sim 0.12$, and then reddened slightly to
$V_{MAC}-R_{MAC} \sim 0.17$ during the shallow dip that preceded outburst
A (see Fig.\ 1). 

There is no unique short period which fits all segments simultaneously. 
It appears that each segment has a quasi-periodicity near 11 days.  Since
it is likely that the 394-day outbursts are due to some disturbance in the
equatorial disk during each orbit, then how would this be manifest after
the outburst?  We note that $\sim$11 days is about the timescale for the
rotation of the outer portion of a 10R$_{\star}$ disk.  Hence, the period
found following an outburst may be due to the rotation of the excited
region where the neutron star penetrated the disk and our changing view
of it.  Variations of $\sim$5\% in the disk radius, and hence its outer 
rotation period, could easily be expected.  Longer periods would indicate a 
greater equatorial disk radius. 

Because the MACHO data are taken at approximately one day intervals, we
cannot rule out that the true periodicity is at the shorter, alias period 
near 1.1 days (labeled `1$-$f' in Figure 4)).  Percy et al.\ (2002) have 
shown that some Be stars exhibit low-amplitude variations with timescales 
of the order of 0.3 to 2 days due to nonradial pulsations or rotation.  
However, this alternate interpretation seems unlikely since such a period 
would not be expected to vary substantially from one segment to another.  
It is more reasonable to associate variations in the periodicity with 
changes in the disk, and hence the $\sim$11-day period is the primary one. 

\section{Summary} 

We have shown that AX~J0049.4$-$7323 exhibits small, recurrent optical 
outbursts with a period of $\sim$394 days.  This period is identified with 
the orbital period of the system and may be caused by the interaction of the
neutron star with the equatorial disk of the Be star during periastron
passage.  During the outbursts, the system's color becomes redder,
consistent with a brightening in the disk. 

In addition, between outbursts the system displays color-independent,
quasi-periodic variability, with a timescale of $\sim$11 days.  We
tentatively suggest that this behavior may result from the observer's
changing view of the disturbed region in the equatorial disk as it rotates
around the Be star. 

\acknowledgments 
We thank the referee, Dr.\ Malcolm Coe, for a helpful report.

\clearpage

\begin{figure}
\caption{Longterm light curves of AX~J0049.4$-$7323.  The top two panels show
MACHO $V_{MAC}$ and $R_{MAC}$ data while the bottom panel displays the 
OGLE-II $I$ light curve.  The outbursts, labeled A through F, occur every 
$\sim$394 days. The segments between outbursts, labeled `Seg 1' through 
`Seg 6', have also been analyzed for periodic behavior.  All segments show 
quasi-periodic behavior with periods of about 11 days. } 
\end{figure}

\begin{figure}
\caption{Outburst `F' ($\sim$ MJD 51400 through 51420), as observed in
$V_{MAC}$, $R_{MAC}$, and $I$. Note that the amplitude of the outburst is
greater at longer wavelengths indicating the system reddens during the
outbursts.  This is also seen in the bottom panel where the color
$V_{MAC}-R_{MAC}$ is plotted. } 
\end{figure}

\begin{figure}
\caption{$R_{MAC}$ light curves and $V_{MAC}-R_{MAC}$ color curves for 
individual outbursts B through E.  Both the duration and amplitude of the 
outbursts are variable, although the general shape is similar, with a rapid 
rise and slow decline.  Note that the color reddens slightly during most 
outbursts.} 
\end{figure}

\begin{figure}
\caption{Sample periodograms of detrended MACHO photometry.  The top
panel shows the periodogram of $R^\ast$ data from Segment 2 covering a wide 
range of frequencies.  The strongest peak (labeled `f') is at P$\sim$11 days, 
but several aliases are present due to the daily sampling interval.  The 
bottom panel shows periodograms for Segments 2, 3, 4, and 5, expanded 
around the primary peaks.  While each segment clearly contains periodic 
behavior near P$\sim$11 days, the periods found for individual segments are 
not exactly the same.}
\end{figure}

\begin{figure}
\caption{ $R^\ast$ light curves for Segments 2 and 3, each folded on its
own best period.}
\end{figure}


\clearpage 

\begin{references}

\reference{} Alcock, C., et al. 1999, PASP, 111, 1539

\reference{} Alcock, C., et al. 2001, \mnras, 321, 678

\reference{} Charles, P.A., et al. 1983, MNRAS, 202, 657

\reference{} Coe, M.J., Negueruela, I., Buckley, D.A.H., Haigh, N.J., \& 
Laycock, S.G.T. 2001, \mnras, 324, 623

\reference{} Corbet, R.H.D. 1984, A\&A, 141, 91

\reference{} Edge, W.R.T., \& Coe, M.J. 2003, MNRAS, 338, 428

\reference{} Haberl, F., \& Sasaki, M. 2000, A\&A, 359, 573

\reference{} Horne, J.H., \& Baliunas, S.L. 1986, \apj, 302, 757

\reference{} Hutchings, J.B., Crampton, D., Cowley, A.P., \&
Olszewski, E. 1985, PASP, 97, 418

\reference{} in't Zand, J.J.M., Swank, J., Corbet, R.H.D., \& Markwardt,
C.B. 2001, A\&A, 380, L26 

\reference{} McGowan, K.E., \& Charles, P.A. 2002, MNRAS, 335, 941

\reference{} Percy, J.R., Hosick, J., Kincaide, H., \& Pang, C. 2002, PASP, 
114, 551

\reference{} Schmidtke, P.C., Cowley, A.P., \& Levenson, L. 2003, AJ, 126, 1017

\reference{} Waters, L.B.F.M., \& van Kerkwijk, M.H. 1989, A\&A, 223, 196

\reference{} Yokogawa, J., Imanishi, K., Ueno, Masaru, \& Koyama, K. 2000,
PASJ, 52, L73 

\reference{} Yokogawa, J., Imanishi, K., Tsujimoto, M., \&  Koyama, K.
2003, PASJ, 55, 161 

\end{references}
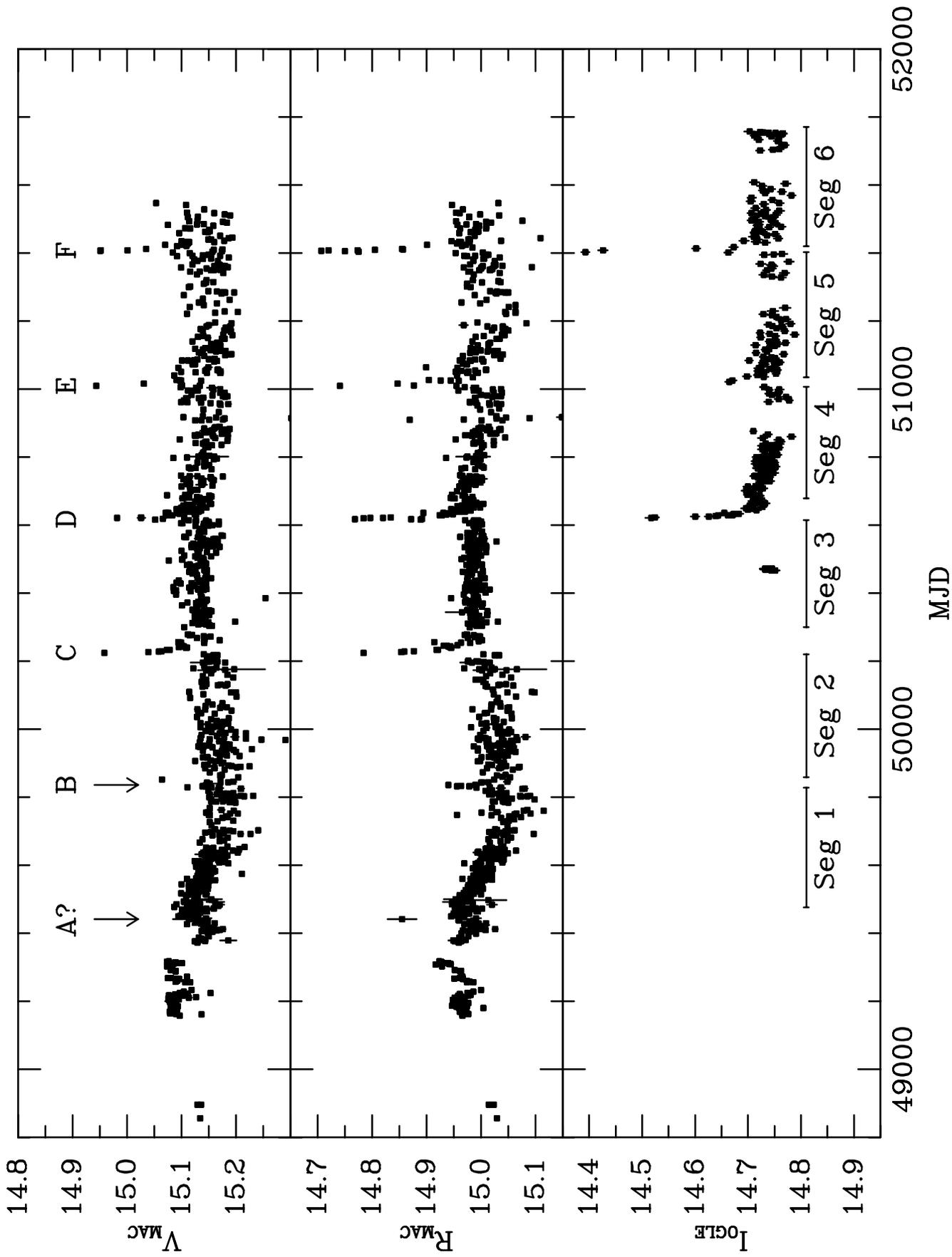
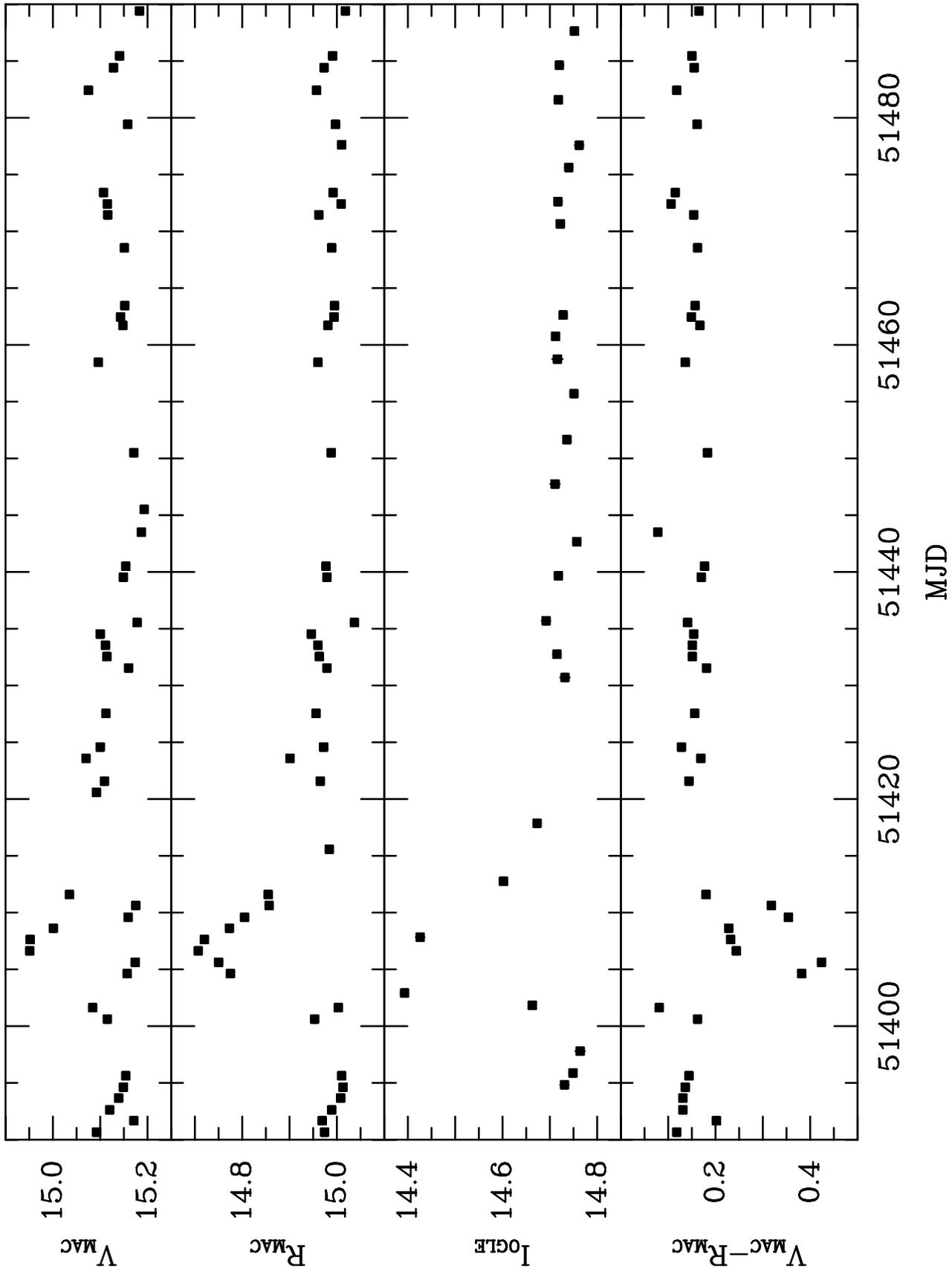
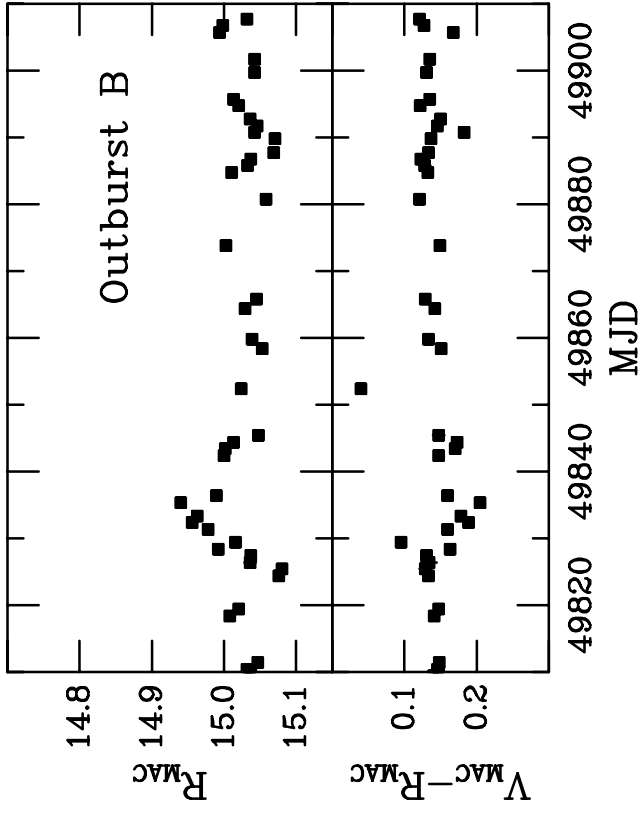
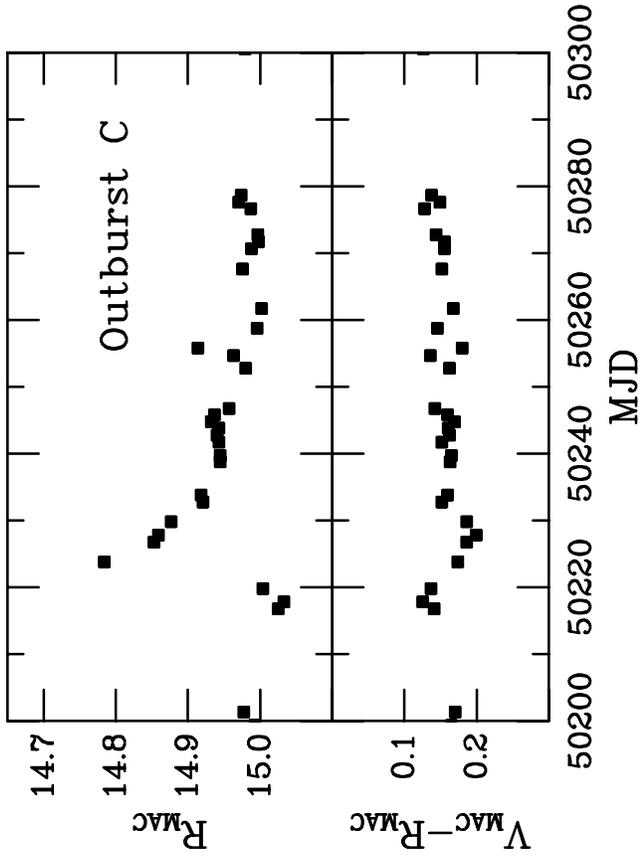
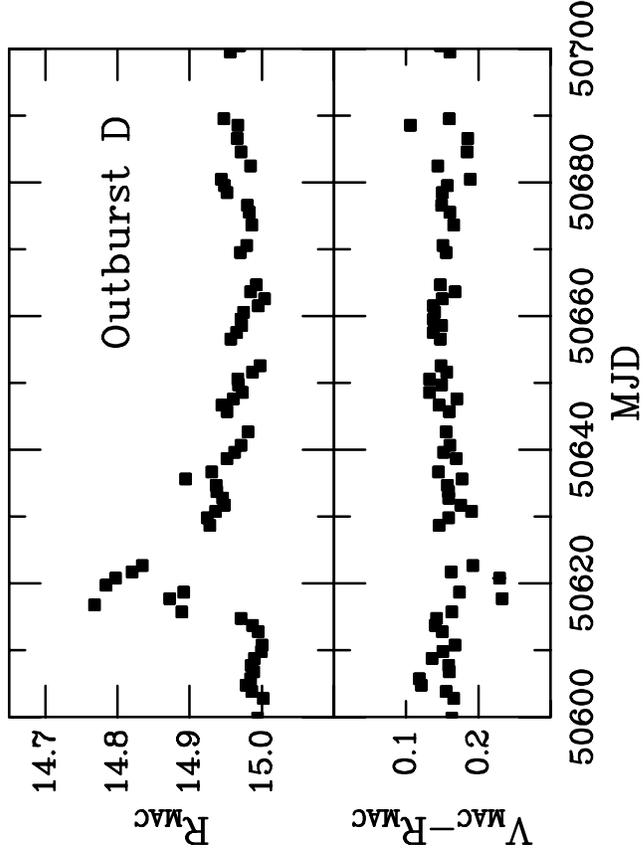
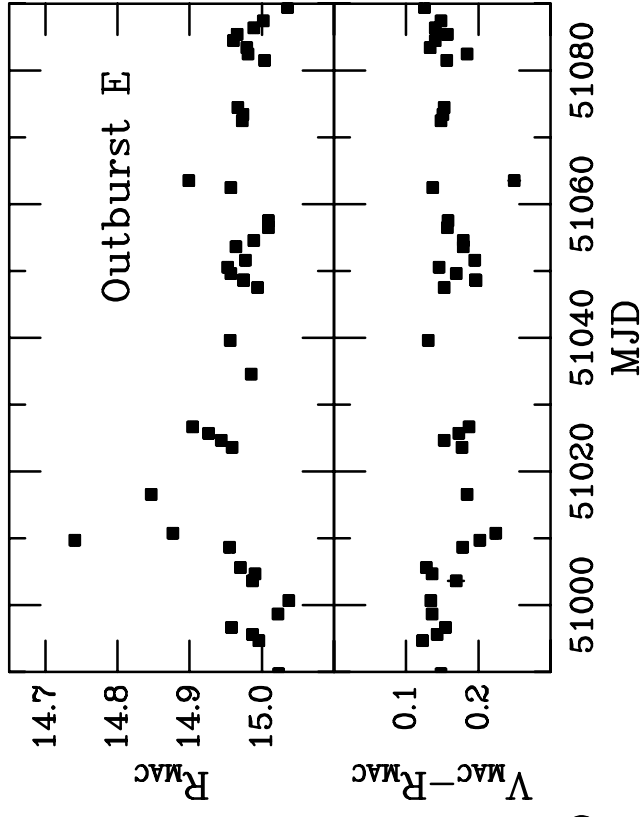
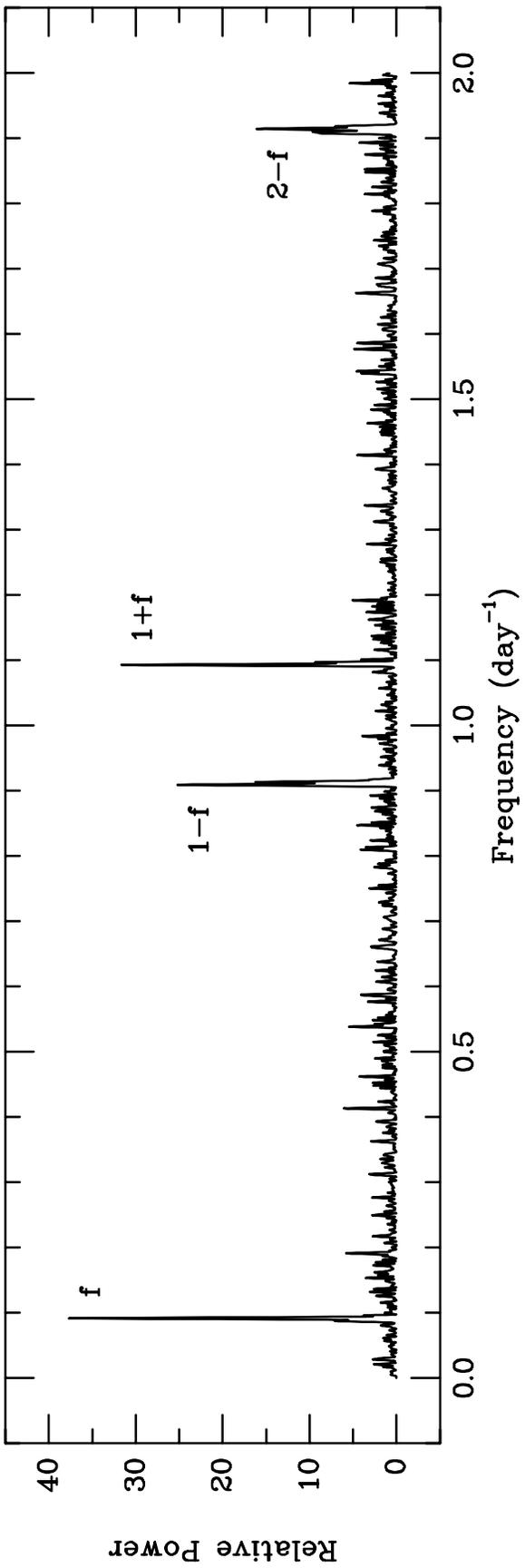
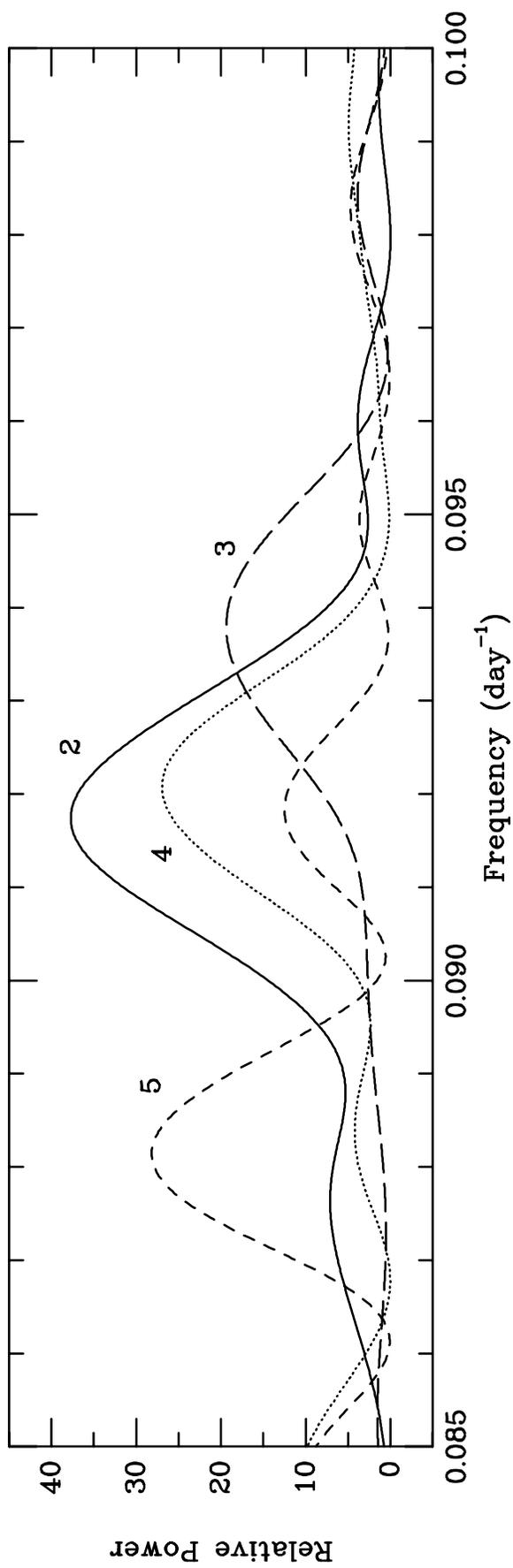
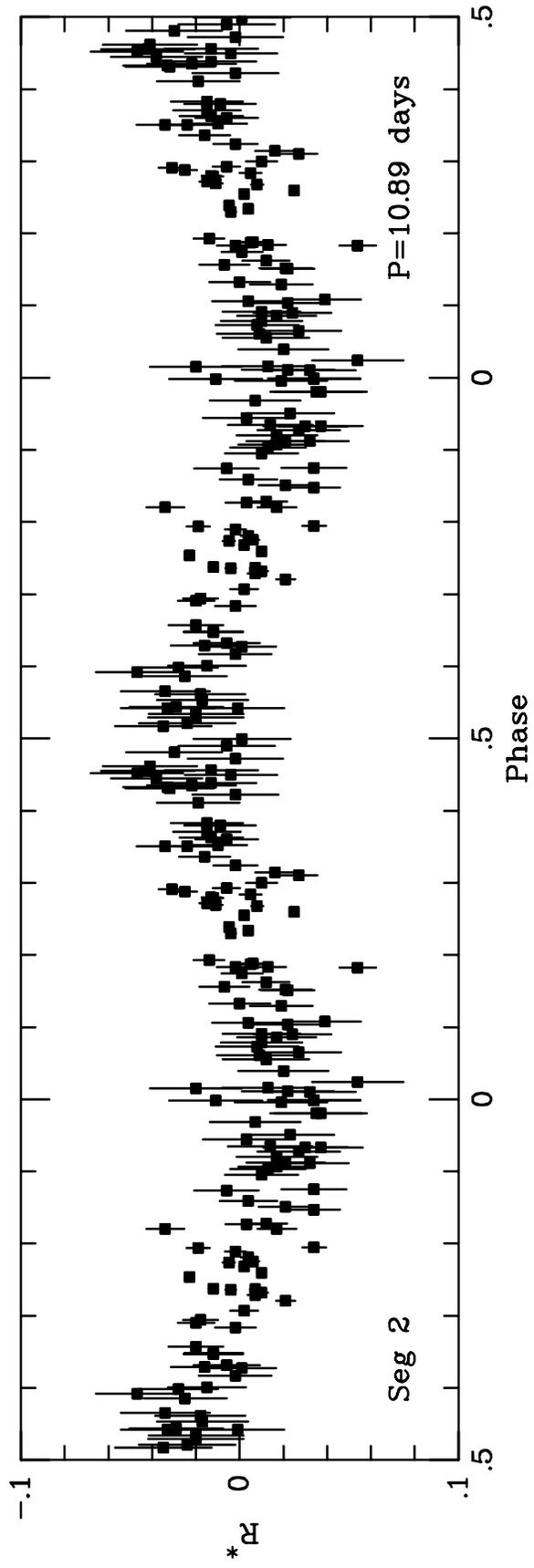
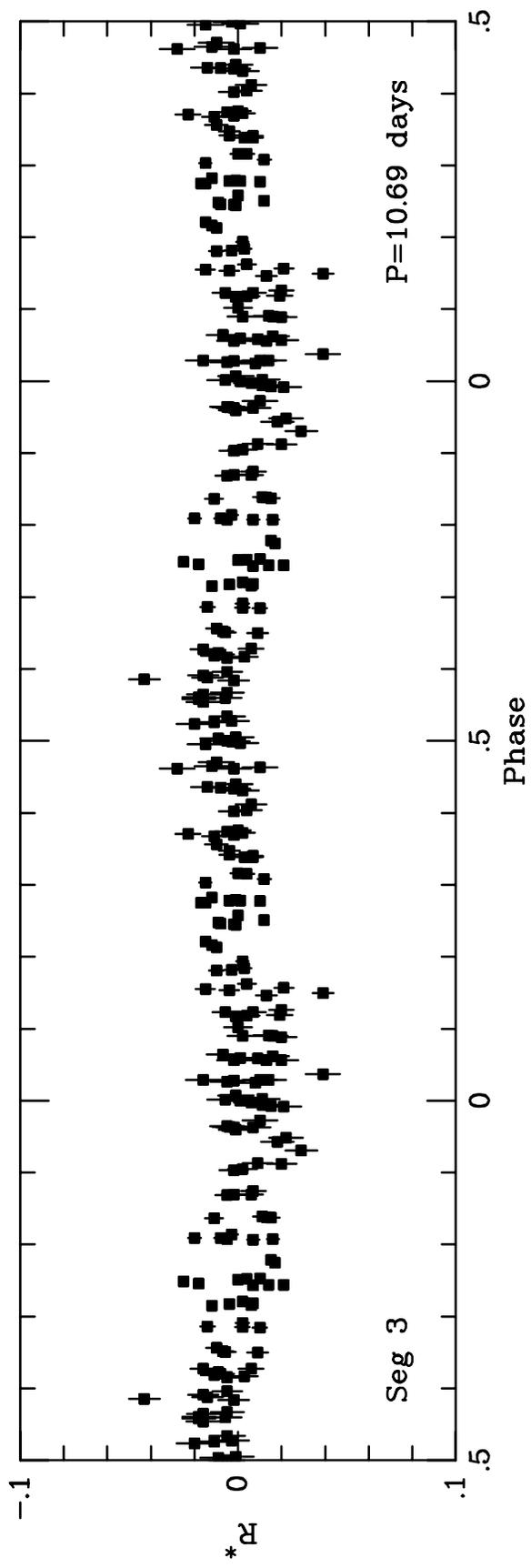
\end{document}